\documentstyle[12pt,epsfig]{article}
\begin{document}
\title{Self-consistent Ornstein-Zernike approximation for three-dimensional
spins}
\date{}
\author{D. Pini \\
{\em Istituto Nazionale per la Fisica della Materia e} \\
{\em Universit\`a degli Studi di Milano, via Celoria 16, 20133 Milano, Italy}
\and J. S. H\o ye \\
{\em Institutt for Fysikk, NTNU,} \\
{\em N 7491 Trondheim, Norway.}
\and G. Stell \\
{\em Department of Chemistry, State University of New York at Stony Brook,} \\
{\em Stony Brook, NY 11794-3400, U.S.A.}}

\maketitle

\begin{abstract}
An Ornstein-Zernike approximation for the two-body correlation function
embodying thermodynamic consistency is applied to a system of classical
Heisenberg spins on a three-dimensional lattice. The consistency condition
determined in a previous work is supplemented by introducing a simplified
expression for the mean-square fluctuations of the spin on each lattice site.
The thermodynamics and the correlations obtained by this closure are
then compared
with approximants based on extrapolation of series expansions and with
Monte Carlo simulations.
The comparison reveals that many properties of the model,
including the critical temperature, are very well reproduced by this
simple version of the theory, but that it shows substantial quantitative
error in the critical region, both above the critical temperature
and with respect to its rendering of the spontaneous magnetization curve.
A less simple but conceptually more satisfactory version of the SCOZA is then
developed, but not solved, in which the effects of transverse correlations
on the longitudinal susceptibility is included, yielding a more complete
and accurate description of the spin-wave properties of the model.

\end{abstract}

\section{Introduction}

Several approaches in the context of liquid-state theory hinge on the
Ornstein-Zernike (OZ) equation that relates the two-body distribution
function $g({\bf r})$ and the direct correlation function $c({\bf r})$.
Once an approximate expression for
$c({\bf r})$ is inserted into this equation, a closed theory for the
thermodynamics and the correlations is obtained.

While the dependence of $c({\bf r})$ on the thermodynamic variables
is fed directly into the expression of $c({\bf r})$ itself in the commonly
used closures to the OZ equation,
another method for obtaining a closed theory amounts to requiring that
$c({\bf r})$ depends on the density and on the temperature of the system
in such a way that consistency between different paths to the thermodynamics
is obtained. Recent results on simple lattice gases~\cite{ising} and
model fluids~\cite{yuk} have shown that such a {\em self-consistent
Ornstein-Zernike approximation} (SCOZA) is able to yield very good results
for both the overall thermodynamics and correlations and for the critical
point and the coexistence curve.

The idea of using thermodynamic consistency as a constraint to close the OZ
equation is relevant not only to simple fluids. In particular, the theory has
already been formulated also for the $D$-vector model~\cite{dvec}, i.e.,
a system of classical $D$-dimensional spins with nearest-neighbor interaction.
In the present paper the $D$-vector model SCOZA is further investigated
and the simplest version of the SCOZA that we have developed is solved
numerically in the
case of three-dimensional spins (classical Heisenberg model) on a
three-dimensional lattice, and the results are compared with
Monte Carlo simulations~\cite{binder,holm,landau} and Pad\`e approximants
based on high-temperature expansions~\cite{domb,wood,fisher,mckenzie,butera}.
An analysis of what is lacking in this simple application to the $D$-vector
model then guides us in formulating a more sophisticated
version of the theory that takes into account the influence of transverse
correlations on the longitudinal susceptibility and as a result, yields
a more complete and accurate description of the spin-wave properties
of the model.

Besides allowing us to test the SCOZA for a system with a vector order
parameter, we believe that the investigation set forth here is also
interesting in its
own sake. While the main properties of the D-vector model have been
understood for a long time and its critical constants have been determined
with great accuracy in several recent papers both by finite-size scaling
techniques~\cite{landau,holm} and by extrapolation of series
expansions~\cite{butera}, there is nevertheless a paucity of treatments
that yield a reasonable description of the model for $D=3$ (the classical
Heisenberg model) over the whole phase plane, especially compared
to the Ising or $D=1$ case.
The $D$-vector model has recently been studied by cumulant expansion
methods~\cite{cluster1,cluster2}, which can be regarded as a systematic way
to improve
the mean-field approximation. However, results were determined only in zero
magnetic field, and do not include the correlation functions. Moreover,
the predictions for the critical temperature obtained within the currently
adopted
truncation of the cumulant expansion are in only modest agreement with the
expected values~\cite{cluster2}.
It seems therefore interesting to develop
a version of SCOZA for this model, particularly for $D=2$ and $D=3$,
that promises to yield high overall accuracy, especially since our approach
gives both the thermodynamics and the correlations of the system.

A difference that arises in the present implementation of the SCOZA
with respect to the Ising case is that the OZ approximation (i.e., the
assumption that the direct correlation function $c({\bf r})$ has the same
range as the interaction between the spins) and the
short-range constraint on the two-body correlation function which stems
from the fixed length of the spins are not enough to get a closed theory
by means of a single thermodynamic consistency condition.
We need some {\em ansatz} about the mean-square fluctuations of the transverse
and longitudinal spin components.
The initial choice we investigate here is to regard these
fluctuations of the spin on each lattice site as a function of the
magnetization alone, thereby using at every temperature the expression
valid in the high-temperature limit. We find that within this approximation
the error in the critical temperature is at most about $0.5\%$, and the
compressibility, the correlation length, and the correlation functions are
nicely reproduced over a wide range of thermodynamic states. Nevertheless,
the global accuracy of this version of the theory is inferior to that
of the Ising or simple
fluid case. In particular, for both the Ising and the Heisenberg model the
SCOZA gives spherical critical exponents above the critical temperature $T_c$,
but in the Ising case~\cite{ising} the asymptotic critical regime is
detectable only very close to the critical point and the thermodynamics is
well reproduced up to reduced temperatures $t=(T-T_c)/T_c\sim 10^{-2}$. For
the Heisenberg model, on the other hand, the deviations from the correct
critical behavior appear already for $t\sim 10^{-1}$. At smaller values of
$t$, the theory readily reaches the asymptotic spherical regime. Moreover,
while in the Ising-model SCOZA the coexistence curve displays a non-classical
exponent $\beta=7/20$~\cite{ising,crit},
the Heisenberg coexistence curve predicted by the SCOZA instead
is associated with
$\beta=1/2$, and is overall considerably narrower
than the expected one as given by MC simulations~\cite{binder} and Pad\`e
approximant techniques~\cite{wood}. (We find however that it displays the
correct behavior in the low-temperature limit).
In order to probe the sensitiveness of the theory to the approximate
{\em ansatz}
adopted for the mean-square spin fluctuations, we also implemented a different
approximation, which was originally proposed in Ref.~\cite{dvec}. The results
are almost identical in the two cases.
In response to the lack of high quantitative accuracy shown by these simple
versions of the theory we then develop (but do not solve here)
an improved version that incorporates a more complete and accurate description
of spin-wave correlations.

The paper is organized as follows: in Sec.~2 we review the theory already
developed in Ref.~\cite{dvec}, introduce an approximation necessary to
obtain a closed equation, and cast it into a form amenable to numerical
solution, which we obtain in Sec.~3. There we show our results
for the critical point, the thermodynamics, the correlations,
and the phase diagram, and compare them to predictions from MC simulations
and Pad\`e approximants.
We give our improved version of the theory in Sec.~4, in which we include
the effects of transverse correlations on the longitudinal susceptibility.
Our findings are briefly summarized in Sec.~5.

\section{Theory}

We consider a system of $D$-dimensional vectors
${\bf s}_{\bf r}$ of unit length on a three-dimensional lattice,
${\bf r}$ being the position
of a generic lattice site, interacting via a nearest-neighbor, ferromagnetic
coupling. Here we are interested in the classical analog of the Heisenberg
model, and therefore will deal with the case $D=3$. The hamiltonian of the
system is
\begin{equation}
{\cal H}=-J\sum_{\langle{\bf r}, {\bf r}'\rangle}
{\bf s}_{\bf r}\!\cdot \!{\bf s}_{{\bf r}'} -
{\bf H}\cdot \sum_{\bf r}{\bf s}_{\bf r} \, ,
\label{heis}
\end{equation}
where the sum over ${\bf r}$ and ${\bf r}'$ extends over all
nearest-neighbor couples, ${\bf H}$ is the external magnetic field, and $J>0$
is the coupling constant, which will be taken equal to unity in the following.
The SCOZA for this model was originally
introduced in Ref.~\cite{dvec}. We begin by reviewing it
in some detail.

The reduced longitudinal and transverse isothermal susceptibilities
$\chi_{\parallel}$, $\chi_{\perp}$ are defined by:
\begin{eqnarray}
\frac{1}{\chi_{\parallel}} & = & \left .\left( \frac{\partial
\beta H_{\parallel}}
{\partial m_{\parallel}}\right)_{\! \beta} \right|_{m_{\parallel}=m}  \, ,
\label{chi1par} \\
\frac{1}{\chi_{\perp}} & = & \left .\left( \frac{\partial \beta H_{\perp}}
{\partial m_{\perp}}\right)_{\! \beta} \right|_{m_{\perp=0}} \, ,
\label{chi1perp}
\end{eqnarray}
where $m$ is the magnetization per spin and $\beta=1/k_{\rm B}T$,
$T$ being the absolute temperature, and $k_{\rm B}$ the Boltzmann's constant.
(We remark that the definitions of $\chi_{\parallel}$, $\chi_{\perp}$
used here differ from those in Ref.~\cite{dvec} by a factor $\beta$).
Here the symbols $\parallel$ and $\perp$ refer respectively to the direction
along the magnetic field ${\bf H}$, and to any of the $D-1$ directions of the
coordinate axes lying in the plane perpendicular to ${\bf H}$. Because of the
spherical symmetry of the system, we find
\begin{eqnarray}
\frac{1}{\chi_{\parallel}} & = & \left(\frac{\partial \beta \! H}
{\partial m}\right )_{\! \beta}=\left(\frac{\partial^2 \beta a}
{\partial m^2}\right )_{\! \beta} \, ,
\label{chi2par} \\
\frac{1}{\chi_{\perp}} & = & \frac{\beta H}{m} =
\frac{1}{m}\left (\frac{\partial \beta a}
{\partial m}\right )_{\! \beta} \, ,
\label{chi2perp}
\end{eqnarray}
where $a=a(m,\beta)$ is the Helmholtz free energy per spin. Let us note that
Eqs.~(\ref{chi2par}),~(\ref{chi2perp}) imply
\begin{equation}
\frac{1}{\chi_{\parallel}}=\frac{\partial}{\partial m}
\left(\frac{m}{\chi_{\perp}}\right )
\, .
\label{sum}
\end{equation}
The internal energy per spin $u$ is given by
\begin{equation}
u=\left(\frac{\partial \beta a}{\partial \beta}\right )_{\! m} \, .
\label{u}
\end{equation}
We now introduce the two-spin correlation function matrix
${\bf \Gamma}({\bf r})$ with elements
\begin{equation}
\Gamma_{ij}({\bf r})=
\langle s^{i}_{\bf r} s^{j}_{\bf 0}\rangle -
\langle s^{i}_{\bf r}\rangle \,
\langle s^{j}_{\bf 0}\rangle \, ,
\label{gamma}
\end{equation}
where the indexes $i$, $j$ refer to two generic components
of the vectors ${\bf s}_{\bf r}$, ${\bf s}_{\bf 0}$.
Isotropy implies that ${\bf \Gamma}$ can be put into
diagonal form by choosing the direction of the magnetic field ${\bf H}$ as one
of the coordinate axis. In particular, we have
\begin{eqnarray}
\Gamma_{\parallel}({\bf r}) & = & \langle s^{\parallel}_{\bf r}
s^{\parallel}_{\bf 0} \rangle -m^2 \, ,
\label{gpar} \\
\Gamma_{\perp}({\bf r}) & = & \langle s^{\perp}_{\bf r}
s^{\perp}_{\bf 0} \rangle \, .
\label{gperp}
\end{eqnarray}
We recall that both the internal energy $u$ and the susceptibilities
$\chi_{\parallel}$, $\chi_{\perp}$ can be expressed by the correlation
functions
$\Gamma_{\parallel}$, $\Gamma_{\perp}$. Specifically, we have for the internal
energy
\begin{equation}
u=-\frac{q}{2}\left[m^2+\Gamma_{\parallel}(\mbox{\boldmath $\lambda$})+
d\, \Gamma_{\perp}(\mbox{\boldmath $\lambda$})\right] \, .
\label{ucorr}
\end{equation}
Here {\boldmath $\lambda$} is a vector joining a given lattice site with
any of its
nearest neighbors, $q$ is the number of nearest neighbors, and $d=D-1$.
(In Ref.~\cite{dvec} the factor $q$ is not used. This corresponds to setting
$Jq=1$). The susceptibilities are given by
\begin{eqnarray}
\chi_{\parallel} & = & \hat{\Gamma}_{\parallel}(k=0)  \, ,
\label{chiparsum}   \\
\chi_{\perp} & = & \hat{\Gamma}_{\perp}(k=0) \, ,
\label{chiperpsum}
\end{eqnarray}
where $\hat{\Gamma}_{\parallel}({\bf k})$,
$\hat{\Gamma}_{\perp}({\bf k})$ are the Fourier transforms of
$\Gamma_{\parallel}({\bf r})$, $\Gamma_{\perp}({\bf r})$. Since the length of
each spin is equal to unity, we have for every site
${\bf r}$
\begin{equation}
\langle \, (s_{\bf r}^{\parallel})^2\, \rangle +
d\, \langle\, (s_{\bf r}^{\perp})^2\, \rangle = 1 \, .
\label{norm}
\end{equation}
If we set
\begin{eqnarray}
& & R_{\parallel} \equiv \Gamma_{\parallel}(r\!=\!0) \, ,
\label{rpar} \\
& & R_{\perp} \equiv \Gamma_{\perp}(r\!=\!0) \, ,
\label{rperp}
\end{eqnarray}
we find from Eqs.~(\ref{gpar}),~(\ref{gperp}) that the condition~(\ref{norm})
becomes
\begin{equation}
R_{\parallel}+d\, R_{\perp}=1-m^2 \, .
\label{core}
\end{equation}
Let us now consider the direct correlation function $c({\bf r})$
whose Fourier transform $\hat{c}({\bf k})$ is related to
$\hat{\Gamma}({\bf k})$ by
\begin{equation}
\Gamma({\bf k})=\frac{R}{1-R\, \hat{c}({\bf k})} \, ,
\label{oz}
\end{equation}
where we have understood the
subscripts $\parallel$, $\perp$ in both
$\hat{\Gamma}({\bf k})$ and $\hat{c}({\bf k})$. In order to simplify the
notation, we will adopt this same convention in the remaining of the Section
whenever a relation applies to both the longitudinal and the transverse
components of the correlation functions. Eq.~(\ref{oz}) corresponds to the
OZ equation commonly used in liquid-state
theory~\cite{hansen}. We note that, since the matrix ${\bf \Gamma}({\bf r})$
is diagonal, Eq.~(\ref{oz}) is a purely scalar relation. Here we use
the {\em ansatz} also known as OZ
approximation, namely we assume that $c({\bf r})$ has the same range as the
interaction between the spins. For the model in study we have then
\begin{equation}
c({\bf r})=\left\{
\begin{array}{ll}
c_0 & {\bf r}=0  \, ,\\
c_1 & {\bf r}=\mbox{\boldmath $\lambda$}  \, ,\\
0   & {\rm otherwise} \, .
\end{array}
\right.
\label{c}
\end{equation}
Eq.~(\ref{c}) is readily written in Fourier space:
\begin{equation}
\hat{c}({\bf k})=c_{0}+qc_{1}\gamma({\bf k})  \, ,
\label{ck}
\end{equation}
where the nearest-neighbor Fourier transform $\gamma({\bf k})$ is given by
\begin{equation}
\gamma({\bf k})=\frac{1}{q}\sum_{{\bf\lambda}={\rm n.n.}}
{\rm e}^{-i{\bf k}\cdot {\bf\lambda}}  \, ,
\label{nn}
\end{equation}
the sum running over all the nearest neighbors of a given lattice site.
We can now use the OZ equation~(\ref{oz}) to express
$\Gamma({\bf r})$ via $\hat{c}({\bf k})$. Let us introduce the
quantity $z$ given by
\begin{equation}
z=\frac{qRc_1}{1-Rc_0}
\label{z}
\end{equation}
and the lattice Green's function
\begin{equation}
P(z)=\int_{\cal B} \!\frac{d^3{\bf k}}{(2\pi)^3}
\, \frac{1}{1-z\gamma({\bf k})} \, ,
\label{green}
\end{equation}
where ${\cal B}$ is the first Brillouin zone of the lattice.
The function $P(z)$ has been widely studied in the
literature~\cite{giap,joyce}. For the simple cubic (SC), body-centered cubic
(BCC), and face-centered cubic (FCC) lattices considered here it can be
evaluated in terms of elliptic integrals~\cite{joyce}.
>From Eqs.~(\ref{rpar})-(\ref{green}) we readily find
\begin{eqnarray}
& & c_0  = \frac{1-P(z)}{R}  \, ,
\label{c0}  \\
& & c_1  =  \frac{zP(z)}{qR} \, ,
\label{c1}  \\
& &\Gamma(\mbox{\boldmath $\lambda$})  =  -R\, \frac{1-P(z)}{zP(z)} \, .
\label{g1}
\end{eqnarray}
The susceptibilities and the internal energy can then be expressed
in terms of $z$ and $R$. From Eqs.~(\ref{chiparsum}), (\ref{chiperpsum}),
(\ref{oz}), (\ref{c0}), (\ref{c1}) we get
\begin{equation}
\frac{1}{\chi}=\frac{(1-z)P(z)}{R}  \, ,
\label{chi}
\end{equation}
while Eqs.~(\ref{ucorr}) and (\ref{g1}) yield
\begin{equation}
u=\frac{q}{2}\left[-m^2+R_{\parallel} \,
\frac{1-P(z_{\parallel})}{z_{\parallel}P(z_{\parallel})}+d\, R_{\perp}
\frac{1-P(z_{\perp})}{z_{\perp}P(z_{\perp})} \right] \, .
\label{uz}
\end{equation}
It appears then that within the OZ approximation the thermodynamics
and the correlations are determined by the quantities $z_{\parallel}$,
$z_{\perp}$, $R_{\parallel}$, $R_{\perp}$. These quantities are not all
independent: in fact, they must be chosen in such a way that the conditions
(\ref{sum}) and (\ref{core}) are satisfied. As noted above, the former is a
consequence of the isotropy of the system, while the latter stems from the
requirement of each spin having unit length.
The other constraint we want to be satisfied is that of consistency between
different routes to the thermodynamics which is peculiar to the SCOZA.
According to the energy route, the thermodynamics is determined from the
internal energy of the system as given by Eq.~(\ref{ucorr}). On the other
hand, knowledge of the correlations
may be used to obtain the thermodynamics also via Eq.~(\ref{chiparsum}) or
(\ref{chiperpsum}) for the susceptibilities. This is the analog of the
compressibility route in fluid systems. While in a hypotetical exact
treatment both routes are obviously bound to yield the the same result,
this is general no
longer true as soon as approximations are introduced. As is readily found
from Eqs.~(\ref{chi2par}), (\ref{chi2perp}), (\ref{u}), the requirement
that the susceptibilities $\chi_{\parallel}$, $\chi_{\perp}$ and the internal
energy $u$ come from a unique Helmholtz free energy is embodied in the
relations
\begin{eqnarray}
\frac{\partial}{\partial \beta}\left(\frac{1}{\chi_{\parallel}}\right)
& = & \frac{\partial^2 u}{\partial m^2} \, ,
\label{consist1}    \\
\frac{\partial}{\partial \beta}\left(\frac{1}{\chi_{\perp}}\right)
& = & \frac{1}{m}\, \frac{\partial u}{\partial m} \, .
\label{consist2}
\end{eqnarray}
We note that these expression are not independent as they are related by
Eq.~(\ref{sum}). Substitution of Eqs.~(\ref{chi}), (\ref{uz}) into
Eq.~(\ref{consist2}) yields the thermodynamic consistency condition
\begin{equation}
\frac{1}{q}\, \frac{\partial}{\partial \beta}
\left[\frac{(1-z_{\perp})P(z_{\perp})}{R_{\perp}}\right] =
-1+\frac{1}{2m}\, \frac{\partial}{\partial m}
\left[R_{\parallel}\, \frac{1-P(z_{\parallel})}
{z_{\parallel}P(z_{\parallel})} +
d\, R_{\perp} \frac{1-P(z_{\perp})}{z_{\perp}P(z_{\perp})}\right]  \, ,
\label{consist}
\end{equation}
and there will be a corresponding equation obtained from Eq.~(\ref{consist1}).
Eqs.~(\ref{sum}), (\ref{core}), and (\ref{consist}) are a set of
three independent conditions for $z_{\parallel}$, $z_{\perp}$,
$R_{\parallel}$, $R_{\perp}$. As we have four unknown quantities,
this is clearly not enough to obtain a closed set of equations. We then
need one more equation, which will be introduced as a further approximate
ansatz besides the OZ approximation~(\ref{c}). This is different from what
is found in the Ising model or $D=1$ SCOZA, in which the OZ ansatz is the sole
approximation introduced in the theory. The closure relation suggested
in Ref.~\cite{dvec} was
\begin{equation}
c_{1}^{\parallel}=c_{1}^{\perp}
\label{closure1}
\end{equation}
i.e., the nearest-neighbor contributions to the direct correlation function in
the longitudinal and in the transverse directions are assumed
to coincide. Eq.~(\ref{c1}) then gives
\begin{equation}
R_{\perp}=R_{\parallel}\frac{z_{\perp}P(z_{\perp})}
{z_{\parallel}P(z_{\parallel})}
\, .
\label{rz}
\end{equation}
On the other hand, Eq.~(\ref{chi}) shows that a divergence of the longitudinal
or transverse susceptibility corresponds respectively to $z_{\parallel}=1$
or $z_{\perp}=1$. As is well known~\cite{spin}, in the $D$-vector model
on a three-dimensional lattice with $D\geq 2$ both susceptibilities are
expected to diverge on the spontaneous magnetization curve. {\em If}
this is actually the case also in the present approximation, then
Eq.~(\ref{rz}) would imply that on the
spontaneous magnetization curve the longitudinal and transverse mean-square
deviations of the spin coincide. This seems quite unphysical, especially
at low temperature, when one expects the behavior of the system to be
governed by small transverse oscillations around the direction of the average
magnetization $m$~\cite{ma}, thereby yielding $R_{\parallel}\ll R_{\perp}$.
As a consequence, although we did implement the approximation of
Eq.~(\ref{closure1}) on order to probe the sensitiveness of the SCOZA to the
closure adopted (see Sec.~3), we decided to resort to a different one for most
of the computations performed here. Specifically, we simply disregarded
the dependence of $R_{\parallel}$, $R_{\perp}$ on the temperature of
the system, i.e. we took for these quantities the same espressions one
finds in the high-temperature limit. Since in this limit the spins on
different sites are uncorrelated, it is found from Eqs.~(\ref{chiparsum}),
(\ref{chiperpsum}) that the present approximation amounts to assuming,
both for the parallel and the transverse components of $R$,
\begin{equation}
R=\chi\,(m,\beta\!=\!0) \, .
\label{closure2}
\end{equation}
The reduced susceptibilities in the high-temperature limit are just those
of the non-interacting spin system. They are then straightforwardly
determined from the Langevin function:
\begin{equation}
m_{\rm id}=\frac{1}{\tanh (\beta H)}-\frac{1}{\beta H}  \, ,
\label{langevin}
\end{equation}
where the subscript ``id'' refers to the non-interacting system.
Eq.~(\ref{langevin}) is then inverted to express $R_{\parallel}$,
$R_{\perp}$ as a function of $m$. We observe that in the approximation of
Eq.~(\ref{closure2}) the condition~(\ref{core}) is
trivially satisfied at every temperature. Eqs.~(\ref{sum}) and
(\ref{closure2}) finally allow one to obtain a closed partial differential
equation (PDE) for the quantity $z_{\perp}$ from the thermodynamic
consistency condition~(\ref{consist}). We now perform some manipulations on
this equation in order to cast it into a form
suitable for numerical integration. Let us note that Eqs.~(\ref{c1}),
(\ref{chi}) imply $0\leq z \leq1$. In this interval the function
\begin{equation}
y=(1-z)P(z)
\label{y}
\end{equation}
is invertible. We then introduce its inverse function $z=F(y)$,
$0 \leq y \leq 1$, and we set
\begin{equation}
G(y)\equiv\frac{1-y-F(y)}{yF(y)}=\frac{1-P(z)}{zP(z)} \, ,
\label{g}
\end{equation}
so that Eq.~(\ref{consist}) becomes
\begin{equation}
\frac{1}{qR_{\perp}}\, \frac{\partial y_{\perp}}{\partial \beta}=
-1+\frac{1}{2m}\, \frac{\partial}{\partial m}
\left[R_{\parallel}\, G(y_{\parallel})+d\, R_{\perp}G(y_{\perp})\right]  \, .
\label{consist3}
\end{equation}
We now use Eq.~(\ref{sum}) along with Eqs.~(\ref{chi}), (\ref{y})
to express $y_{\parallel}$ in terms of $y_{\perp}$:
\begin{equation}
y_{\parallel}=R_{\parallel}\, \frac{\partial}{\partial m}
\left(m\frac{y_{\perp}}{R_{\perp}}\right)  \, .
\label{ypar1}
\end{equation}
In the present approximation in which $R_{\parallel}$, $R_{\perp}$ coincide
with the longitudinal and transverse susceptibilities of the non-interacting
system, it is readily found using Eq.~(\ref{sum}) that Eq.~(\ref{ypar1})
becomes
\begin{equation}
y_{\parallel}=y_{\perp}+m\frac{R_{\parallel}}{R_{\perp}} \,
\frac{\partial y_{\perp}}{\partial m}  \, .
\label{ypar}
\end{equation}
Moreover, we find
\begin{equation}
\frac{\partial y_{\parallel}}{\partial m}=
\left(2+\frac{m}{R_{\perp}}{R'}_{\parallel}\right)
\frac{\partial y_{\perp}}{\partial m}+m\frac{R_{\parallel}}{R_{\perp}} \,
\frac{\partial ^2 y_{\perp}}{\partial m^2}  \, ,
\label{dypar}
\end{equation}
where we have indicated with ${R'}_{\parallel}$ the first derivative of
$R_{\parallel}(m)$. Eqs.~(\ref{ypar}), (\ref{dypar}) are then substituted
into Eq.~(\ref{consist3}) to yield
\begin{eqnarray}
\frac{2}{q}\, \frac{\partial y_{\perp}}{\partial \beta} & = &
R_{\parallel}^2{G'}(y_{\parallel})\frac{\partial ^2 y_{\perp}}
{\partial m^2} +
\left\{R_{\perp}\left[2R_{\parallel}{G'}(y_{\parallel})+
dR_{\perp}{G'}(y_{\perp})
\right]+mR_{\parallel}{R'}_{\parallel}{G'}(y_{\parallel})\right\}
\frac{1}{m} \,
\frac{\partial y_{\perp}}{\partial m}  \nonumber \\
& & \mbox{}+ \frac{R_{\perp}}{m}\left[{R'}_{\parallel}G(y_{\parallel})+
d{R'}_{\perp}
G(y_{\perp})-2m\right]  \, .
\label{scoza}
\end{eqnarray}
Since $R_{\parallel}$, $R_{\perp}$ are known functions of $m$ and
$y_{\parallel}$ is given by Eq.~(\ref{ypar}), the above equation is a closed
PDE for $y_{\perp}$ as a function of $\beta$ and $m$. Let us observe that
Eq.~(\ref{y}) and Eq.~(\ref{chi})
imply that both $y_{\parallel}$ and $y_{\perp}$ are even functions
of $m$. As a consequence, the term
$m^{-1}\partial y_{\perp}/\partial m$ which appears in the r.h.s. of
Eq.~(\ref{scoza}) does not imply any singularity at $m=0$, since in this
limit it just gives $\partial^{2}y_{\perp}/\partial m^2$.
The PDE~(\ref{scoza}) is of course non-linear. In particular, both
$y_{\perp}$ and $\partial y_{\perp}/\partial m$ (via Eq.~(\ref{ypar}))
appear as arguments of the functions $G$, $G'$. However,
the dependence on
$\partial^{2}y_{\perp}/\partial m^2$ is linear. When solving Eq.~(\ref{scoza})
numerically, one can then take advantage
of this structure by resorting to some of the finite-difference algorithms
especially devised for {\em quasi-linear} diffusive
PDE's~\cite{ames}. While these algorithms share with fully implicit methods
the unconditional numerical stability, they do not require the solution of a
large system of coupled non-linear equations at each integration step and
are therefore easier to implement.

In order to integrate Eq.~(\ref{scoza}) we also need a set of initial
and boundary conditions. The initial condition is given at $\beta=0$, i.e.
in the high-temperature limit. In this limit all the correlations vanish
except onsite so that, as we pointed out above,
$\chi_{\perp}$ and $R_{\perp}$ coincide. Eq.~(\ref{chi}) implies then
$z_{\perp}=0$, which in turn gives via Eq.~(\ref{y}) the initial condition
\begin{equation}
y_{\perp}(m,\beta\!=\!0)=1 \mbox{\hspace{1cm}{\rm for every $m$.}}
\label{initial}
\end{equation}
The magnetization per spin $m$ varies in the interval
$-1\leq m\leq 1$. However, since $y_{\perp}$ is an even function of $m$, we
can restrict the integration of Eq.~(\ref{scoza}) to the interval
$0\leq m\leq 1$ provided we require $y_{\perp}$ to be indeed
symmetric at $m=0$:
\begin{equation}
\frac{\partial y_{\perp}}{\partial m}(m\!=\!0,\beta)=0
\mbox{\hspace{1cm}{\rm for every $\beta$.}}
\label{bound1}
\end{equation}
As $m$ approaches its limiting value $m=1$, the longitudinal and transverse
susceptibilities $\chi_{\parallel}$ and $\chi_{\perp}$ vanish respectively
as $(1-m)^2$ and $1-m$. In this limit the reciprocal susceptibilities
diverge like those of the non-interacting
system, as the excess contribution stays finite. We have then again
$1/\chi_{\perp}\sim 1/R_{\perp}$ and hence from Eqs.~(\ref{chi}),
(\ref{y})
\begin{equation}
y_{\perp}(m\!=\!1,\beta)=1 \mbox{\hspace{1cm}{\rm for every $\beta$.}}
\label{bound2}
\end{equation}
We note that for the transverse susceptibility to be positive,
$y_{\perp}$ has to be positive as well. The numerical solution of
Eq.~(\ref{scoza}) shows that this condition is always satisfied if the
inverse temperature $\beta$ is smaller than a certain value $\beta_{c}$ such
that $y_{\perp}(m\!=\!0,\beta_{c})=0$. At
$\beta=\beta_{c}$ both the transverse and (from Eq.~(\ref{ypar}))
the longitudinal susceptibilities in zero field diverge. This is the inverse
critical temperature of the system. For $\beta>\beta_{c}$, we find
that the condition $y_{\perp}\geq 0$ is not satisfied on the whole interval
$0\leq m\leq 1$. Instead, a certain temperature-dependent value $m_{S}(\beta)$
exists, such that
$y_{\perp}$ vanishes for $m=m_{S}$ and becomes negative for
$m<m_{S}$. Since from Eqs.~(\ref{chi2perp}), (\ref{chi}), (\ref{y})
$y_{\perp}=0$ implies $H=0$, $m_{S}$ is just the spontaneous
magnetization at the temperature considered. For $m<m_{S}$ not only does the
transverse susceptibility behave unphysically, but the PDE itself becomes
intrinsically unstable and cannot therefore be integrated numerically. Hence,
we excluded the interval $(0,m_{S})$
from the domain of integration. Below the critical temperature the boundary
condition~(\ref{bound1}) is then replaced by
\begin{equation}
y_{\perp}(m\!=\!m_{S},\beta)=0  \, .
\label{bound3}
\end{equation}
The spontaneous magnetization $m_{S}$ is determined by checking the sign
of $y_{\perp}$ as the integration with respect to $\beta$ goes on. As soon
as $y_{\perp}$ becomes negative at a certain $m=m_{0}$, the spontaneous
magnetization curve is widened by setting
$m_{S}=m_{0}+\Delta m$, $\Delta m$ being the step of the magnetization grid.

The numerical integration of Eq.~(\ref{scoza}) has been carried out with an
inverse temperature step $\Delta \beta$ initially set at
$10^{-4}$--$10^{-5}$. This can be further reduced as the critical region
is approached so as to get very close to the critical temperature, and then
gradually expanded back to its initial value. The number of mesh points in
the magnetization grid ranged from $\sim 10^{3}$ to $\sim 10^{4}$.

\section{Results for the simple version}

In order to assess the reliability of the approximate
closure~(\ref{closure2}), we considered first the case of a linear chain of
three-dimensional spins, which can be solved exactly for zero magnetic
field~\cite{onedim}. The SCOZA predictions agree very well with the
exact solution, as can be seen from
Figs.~1 and~2, where the results for the susceptibility and the specific
heat are shown.
We recall that this
one-dimensional model has no phase transition at non-zero temperature.

We then turn to the three-dimensional lattices. In this case the SCOZA does
predict a phase transition at a certain temperature,
which can be determined by locating the divergence of the isothermal
susceptibility in zero field. The critical temperatures predicted by the SCOZA
for the simple cubic (SC), body-centered cubic (BCC), and face-centered cubic
(FCC) lattices are reported in Table~I, together with the results
of extrapolation of high-temperature expansions~\cite{mckenzie,butera}
and finite-size scaling (FSS) analysis performed on Monte Carlo
simulations~\cite{holm,landau}.
The agreement is very good; the error is at most $0.55\%$.
The critical temperature predicted for the SC lattice
by the renormalization-group based quantum hierarchical reference theory
(QHRT) in the classical limit~\cite{gianinetti} is also shown for comparison.
Furthermore, the Table shows the critical internal energy per
spin $u_{c}$. The SCOZA value of $u_{c}$ is actually the same for any theory
where the OZ approximation~(\ref{c}) and
the normalization condition~(\ref{core}) are satisfied. From
Eq.~(\ref{uz}) with $m=H=0$, $z_{\parallel}=z_{\perp}=1$ one has
$u_{c,{\rm SCOZA}}=-q[P(1)-1]/[2P(1)]$. We note that this result depends
only on the type of lattice structure, so that for a certain lattice the
SCOZA predicts the same $u_{c}$ for any spin dimensionality. The
agreement of the SCOZA result with the data from high-temperature series
expansions~\cite{fisher} appears to be only modest, and sensibly worse than in
the Ising or $D=1$
case~\cite{ising}. This is somewhat surprising, since the SCOZA is known
to be exact in the $D\rightarrow \infty$ or spherical-model limit~\cite{dvec},
so one might have expected very good agreement
for intermediate $D$ as well. Since the SCOZA prediction for $u_{c}$ is not
affected by the approximate closure relation~(\ref{closure2}), such a behavior
seems to indicate that as the spin dimensionality is increased, the
OZ approximation does not become more and more accurate in a monotonic
fashion, but a ``lowest accuracy'' for some intermediate $D$ might be reached.
In Fig.~3 the reduced isothermal susceptibility in zero field above
the critical temperature $\chi_{\parallel}=\chi_{\perp}$ and the
correlation length $\xi$ (BCC lattice) are compared with the results
from extrapolation of high-temperature expansions~\cite{fisher}.
In both cases the agreement appears to be very good, even in the
near-critical regime when
$\xi \sim 10^{2}$. As stated in the Introduction, our numerical procedure
for the $D$-vector model
SCOZA gives for the critical exponents the spherical model values $\gamma=2$,
$\nu=1$, $\beta=1/2$, $\alpha=-1$, $\eta=0$, where the usual notation for the
critical exponents has been used. In
Fig.~4, the susceptibility in the critical region as a function of the reduced
temperature $t=(T-T_c)/T_c$ has been plotted on a log-log scale.
We have also reported the effective critical exponent
$\gamma_{\rm eff}=-d(\log\chi)/d(\log t)$, which shows that for the Heisenberg
model the onset of the asymptotic critical regime as
given by the SCOZA takes place for $t\simeq 10^{-3}$. This is in sharp
contrast to the Ising case, also shown in the Figure,
whose $\gamma_{\rm eff}$ is affected
by a very strong crossover, remaining very accurate over a quite wide range
of $t$ and saturating to its inaccurate asymptotic value only
for $t\simeq 10^{-5}$.
The specific heat $c_{H}$ in zero
field has been plotted in Fig.~5 together with a closed-form
approximant~\cite{domb}. Because of its mean-spherical critical behavior, the
SCOZA predicts that at the critical point $c_{H}$ will saturate at a finite
value, which for the FCC lattice to which the Figure refers is given by
$c_{\rm peak}/k_{\rm B}= 3.03$.
For the Heisenberg model $c_{H}$ is indeed finite at the critical
point.
The approximant reported in the figure
gives $c_{\rm peak}/k_{\rm B}=10$, but one has to take into account
that this quantity depends very sensitively on the value of the
critical exponent $\alpha$ used in the approximant,
which in Ref.~\cite{domb} was taken
equal to $\alpha=-1/16$. As reported in the same reference, using
$\alpha=-1/8$ shifts the peak to $c_{\rm peak}=5.7$, which is probably
a better approximation of the correct value as this choice of $\alpha$
is much closer to the best currently available estimate
$\alpha\simeq -0.122$~\cite{justin}.
The SCOZA predictions for the magnetic structure factor
$\hat{\Gamma}({\bf k})$ in zero field (BCC lattice) are shown in Fig.~6 at
two different temperatures, corresponding to reduced temperatures
$t\simeq 2.2\cdot 10^{-1}$ and $t\simeq 1.4\cdot 10^{-2}$. The agreement with
the approximant~\cite{fisher} is very good.

Let us now consider the results below the critical temperature.
The spontaneous magnetization curve of the SC lattice is reported in
Fig.~7, together with the results of the QHRT~\cite{gianinetti}
and MC simulations~\cite{binder}.
In Fig.~8 the curve of the FCC lattice is compared to a Pad\`e approximant.
We see that in both cases the agreement is relatively poor, with the SCOZA
results looking rather mean-field like. Actually,
in the temperature range shown in the Figures, the main differences between
the SCOZA and the mean-field phase boundaries can be accounted for by the
shift in the critical temperature, so that they look very close to each other,
provided they are plotted as a function of their respective rescaled
temperatures $T/T_{c}$. The effective exponent $\beta_{\rm eff}$ given by the
SCOZA is plotted
in Fig.~9. The critical behavior of the SCOZA phase boundary for the spin
systems with continuous symmetry considered here is different from that
found in the Ising or simple fluid models. In that case, the SCOZA critical
exponents below $T_{c}$ do not coincide with the spherical ones found above
$T_{c}$, and the exponent $\beta$ is
given by $\beta=7/20=0.35$, which gives a very accurate coexistence curve.
Here instead we find spherical exponents both above and below $T_{c}$.
We have then $\beta=1/2$, as in the mean-field approximation, so that
the spontaneous magnetization curve close to the critical temperature
is considerably narrower than the correct
one, for which $\beta\simeq 0.36$~\cite{holm,landau}.
On the other hand, is seems unlikely that this mean-field like behavior
alone can account for the lack of global accuracy of the SCOZA phase boundary.
In fact the QHRT~\cite{gianinetti} yields for $\beta$ the non-classical value
$\beta=0.41$,
but nevertheless is no more accurate than SCOZA as far as overall agreement
with the MC coexistence curve is concerned. An improved closure of the SCOZA
equation expected to deliver more accurate results for the phase diagram
is discussed in the following Section. Here we observe that, despite
its modest overall accuracy, the SCOZA phase boundary becomes
exact in the low-temperature limit. In this regime one can use spin-wave
theory,
which gives for the spontaneous magnetization~\cite{spin}
\begin{equation}
m_{S}\sim 1-\frac{1}{q}P(1)\, k_{\rm B}T+{\cal O}(T^{2}) \, .
\label{wave}
\end{equation}
The spin-wave result for the spontaneous magnetization is compared
to the SCOZA and to the mean-field predictions in Fig.~10. We see that the
SCOZA predicts both the linear dependence on $T$ for
$T\rightarrow 0$ and the correct slope of the ${\cal O}(T)$ term.
Another point which deserves some attention concerns the behavior of the
susceptibility across the spontaneous magnetization curve. For spin systems
with continuous symmetry, we expect the longitudinal susceptibility to
diverge not only at the critical point, but on the whole coexistence curve.
Within the SCOZA, the coexistence curve should then merge with the spinodal.
This is indeed what happens, as shown in Fig.~11, which reports the SCOZA
inverse longitudinal susceptibility
$1/\chi_{\parallel}$
along two different sub-critical isotherms.
In Fig.~12 several super- and sub- critical
isotherms are compared to MC simulation results~\cite{binder}. The discrepancy
between the SCOZA and the MC data already observed in
the coexistence curve affects the sub-critical isotherms for
$H\rightarrow 0$, especially for $T$ near $T_{c}$. The agreement rapidly
improves as $H$ increases.

In order to assess to which extent the results discussed here depend
on the {\em ansatz}~(\ref{closure2}), we also solved numerically the SCOZA
PDE with a different closure relation, namely that of
Eq.~(\ref{closure1}). We find that the thermodynamics and the phase boundary
are hardly distinguishable from those already shown, the critical temperature
actually getting sligthly worse: for instance,
for the SC lattice one has $\beta_{c}=0.6978$, while the previous
approximation (see Table~I) gave $\beta_{c}=0.6968$, and the ``exact'' result
is~\cite{landau,holm,butera} $\beta_{c}=0.6930$. Moreover, our algorithm
failed to find solutions below a certain
sub-critical temperature. We do not know whether this is a purely
numerical problem, or instead it is the sign of a deeper disease which may
affect the closure~(\ref{closure1}). In this respect, we recall that in such
an approximation (see Sec.~II) the mean-square fluctuations of the spin in
the transverse and in the longitudinal directions do not behave correctly on
the coexistence curve, while the correct behavior for $m=1$ is enforced by
the boundary
condition~(\ref{bound2}). It is then possible that at low
temperatures, as the spontaneous magnetization approaches its limiting
value $m_{S}=1$, the change in behavior across the domain becomes so
abrupt as to prevent the existence of a solution. Since the above
closure does not introduce any improvement with respect to the
approximation~(\ref{closure2}) that we have used so far, we have not
investigated this point any further.

\section{Inclusion of induced spin-wave correlations}

\subsection{New closure}

We now seek to develop an improved version of the SCOZA, guided by the results
for the simple version that we have just summarized.
The most serious deficiency of that version is
its spherical-model like coexistence curve, which considerably differs
from what is expected to be the correct result. Thus we have to look for
a possible leading correction that might remedy this.
What we intend to do is to explicitly take into account the effect
of the transverse correlations on the longitudinal susceptibility.
As we will see below, this will naturally give rise to a diverging
longitudinal susceptibility $\chi_{\parallel}$ on the coexistence curve
by enforcing $y_{\parallel}=0$ whenever $y_{\perp}=0$. On the other hand,
if the effect of transverse correlations is not properly accounted for,
a diverging $\chi_{\parallel}$ at coexistence can be obtained only by heavily
distorting the longitudinal part of the correlation function,
with a corresponding infuence upon the internal energy and thus
the SCOZA results. This is most probably what happens in the simple version
that we have considered so far.
For magnetization $m\rightarrow 1$ one approaches Gaussian-model behavior
and these induced correlations can thus be obtained exactly by
a spin-wave calculation.
For general $m$, we can turn to the analysis based on $\gamma$-ordering
in order to obtain their form, where $\gamma$ is the inverse range
of interaction.

The induced correlations become long-ranged near the critical point
and at phase coexistence. Thus their major influence will be upon
the longitudinal susceptibility $\chi_{\parallel}$. Representing
their contribution integrated over space by $\Delta$
we can write, for the longitudinal susceptibility,
\begin{equation}
\chi_{\parallel}=\frac{R_{\parallel}}{y_{\parallel}}=
\frac{R_{\parallel}}{y_{l}}+\Delta \, ,
\label{chimod}
\end{equation}
where the new quantity $y_{l}$ represents the contribution
of the MSA-like part of the correlation function.
Solving with respect to $y_{l}$ one then has
\begin{equation}
y_{l}=\frac{y_{\parallel}\, R_{\parallel}}{R_{\parallel}-y_{\parallel}\Delta}
\, ,
\label{ymod}
\end{equation}
and there will be a quantity $z_{l}$ related to $y_{l}$ by Eq.~(\ref{y}).
The direct contribution these induced correlations will make
to the internal energy will be minor and less crucial according
to $\gamma$-ordering, so we neglect it here for simplicity.
The quantity $y_{l}$ is then used instead of $y_{\parallel}$
in Eq.~(\ref{consist3}) to evaluate the longitudinal part
of the internal energy. This requires having an expression for $\Delta$.
The analysis is most conveniently given in the diagrammatic language
developed~\cite{hemmer} to describe the $\gamma$-ordering used by H\o ye
and Stell. In that language, the correlation functions are built up of chain
bonds (representing chains of pair interactions) and hypervertices
(representing the short-range components of the correlation functions).
The leading contribution beyond the MSA to the pair correlation function
$\Gamma({\bf r})$ of a fluid system has the diagrammatic form

\begin{figure}[htbp]
\begin{center}
\psfig{file=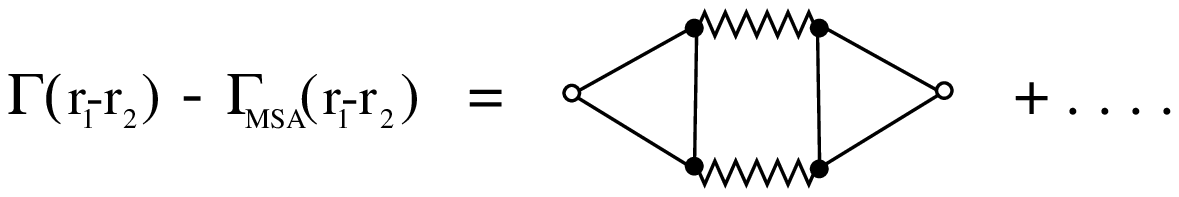}
\end{center}
\end{figure}

\noindent
where the triangles represent the three-body correlation function
of the reference system with purely short-range interactions 
(including self-correlations where two or all three of the particles
can be identical),
the white circles (or ``root points'') represent the coordinates
${\bf r}_{1}$, ${\bf r}_{2}$, and the full circles represent coordinates
that are integrated over.
The zigzag chain bond $C({\bf r})$ (not to be confused with the direct
correlation function $c({\bf r})$) is defined by

\begin{figure}[htbp]
\psfig{file=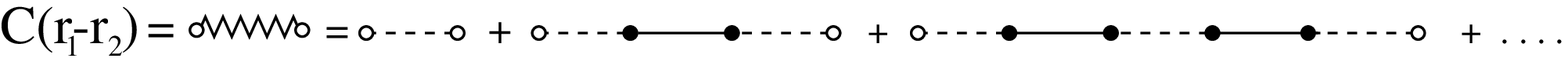,width=16cm}
\label{fig:diag2}
\end{figure}

\noindent
where the solid and dashed lines represent respectively the two-body
correlation function of the reference system $\Gamma_{\!\rm R}({\bf r})$
and the perturbing function $-\beta \varphi({\bf r})$. In Fourier space
one then has
\begin{equation}
\hat{C}({\bf k})=\frac{-\beta\hat{\varphi}({\bf k})}
{1+\beta\hat{\varphi}({\bf k})\hat{\Gamma}_{\!\rm R}({\bf k})} \, .
\label{bond}
\end{equation}
The above result can be generalized to spin systems by utilizing Onsager's
recipe for regarding different orientations as different species of a 
mixture\cite{onsager}.
In particular, the reference system consists of non-interacting spins,
and is mapped into a multicomponent, but simple, hard-core lattice gas whose
$n$-body correlation functions $\Gamma_{\rm R}^{(n)}({\bf r}_{1}\ldots
{\bf r}_{n})$ are non-vanishing only when the coordinates ${\bf r}_{1} \ldots
{\bf r}_{n}$ coincide. In particular, the three-body hypervertex has the form
\begin{equation}
{\bf \Gamma}_{\rm R}^{(3)}({\bf r}_{1},{\bf r}_{2},{\bf r}_{3})=
{\bf b}\,  \delta_{{\bf r}_{1} {\bf r}_{2}} \delta_{{\bf r}_{2} {\bf r}_{3}}
\, .
\label{gamma3}
\end{equation}
In the multicomponent case, the ${\bf \Gamma}_{\rm R}^{(3)}$ as well as
the chain bond should be summed over the various species. One finds
\begin{equation}
\sum_s s\frac{\partial}{\partial \mu_s} = \frac{\partial}{\partial H} \, ,
\end{equation}
where $\mu_s$, the chemical potential has the form $sH + {\it const}$.
Thus the sum on the left-hand side can be replaced by a differentiation
with respect to $H$, which implies that the resulting 
${\bf \Gamma}_{\rm R}^{(3)}$ can again be represented by the 
one-component form Eq.~(51).
For the multicomponent system considered here, ${\bf b}$ is a three-index
tensor with components $b_{ijk}$, where each index can assume $D$ values
corresponding to the longitudinal and each of the $D-1=d$
transverse components of the spin. $b_{ijk}$ is determined by the sum rule
\begin{equation}
b_{ijk}=\frac{1}{N}\!\sum_{{\bf r}_{1},{\bf r}_{2},{\bf r}_{3}} \!\!\!
\Gamma_{\rm R}^{(3)ijk}({\bf r}_{1},{\bf r}_{2},{\bf r}_{3})=
-\left.\frac{\partial^{3} \beta \omega_{\rm R}}{\partial \beta \! H_{i}
\partial \beta H_{j} \partial \beta H_{k}}\right|_{H_{\parallel}=H,
H_{\perp}=0} \, ,
\label{sum3}
\end{equation}
where $N$ is the number of lattice sites and $\omega_{\rm R}$
is the grand potential per site of the reference system. This is the analog
of the compressibility rule (\ref{chiparsum}), (\ref{chiperpsum})
for the two-body correlation function. We recall that for the reference system
the latter gives $\chi_{\parallel}=R_{\parallel}$, $\chi_{\perp}=R_{\perp}$,
where the quantities $R_{\parallel}$, $R_{\perp}$ have been defined
in Eqs.~(\ref{rpar}), (\ref{rperp}). According to Eq.~(\ref{chiparsum}),
the contribution $\Delta$ in Eq.~(\ref{chimod}) is obtained by integrating
over space the leading contribution to
${\bf \Gamma}({\bf r})-{\bf \Gamma}_{\rm MSA}({\bf r})$ shown above,
where each of the white circles corresponds
to a longitudinal root point, and the full circles involve both
a spatial integration and a summation over the component index.
We observe that the matrix of the two-body correlation function
$\Gamma_{\!\rm R}^{ij}$ (see Eqs.~(\ref{gpar}), (\ref{gperp}))
and that of the nearest-neighbor spin-spin interaction
$s_{\bf r}^{i}s_{{\bf r}'}^{j}$ are diagonal. Eq.~(\ref{bond}) then implies
that this is true also for the matrix of the chain bond $C_{ij}({\bf r})$,
i.e., the zigzag bond is non-vanishing only if its endpoints correspond
to the same spin component. It is then readily seen that in performing
the summation over the indexes, the only two terms which contribute
to $\Delta$ are those in which both the three-body hypervertices are either
of the type $b_{\parallel\parallel\parallel}$ or of the type
$b_{\parallel\perp\perp}$. By setting $b_{\parallel\perp\perp}=b$
and applying Eq.~(\ref{sum3}) we find
\begin{equation}
b=\frac{\hspace{0.2cm}\partial R_{\perp}}{\partial \beta \! H}=
R_{\parallel}R'_{\perp}=\frac{R_{\perp}}{m}(R_{\parallel}-R_{\perp}) \, ,
\label{vertex}
\end{equation}
where the prime denotes differentiation with respect to $m$,
and Eq.~(\ref{sum}) has been used.
We note that, since $R_{\parallel}-R_{\perp}={\cal O}(m^{2})$
for $m\rightarrow 0$, $b$ vanishes for $m=0$.
In the limit $m\rightarrow 1$, $b_{\parallel\parallel\parallel}=
R_{\parallel} R'_{\parallel}$ can be disregarded with respect to $b$ as then
one has $b\sim (1-m)^{2}$, $b_{\parallel\parallel\parallel}\sim (1-m)^{3}$
(see the following section). We have then
\begin{equation}
\Delta=\frac{1}{2}d\, b^{2}B \, ,
\label{delta}
\end{equation}
where the factor $1/2$ is the graph symmetry factor with respect
to the interchange of the two chain bonds, $d$ is the number
of the transverse components, and $B$ is given by
\begin{equation}
B=\sum_{\bf r}C^{2}_{\perp}({\bf r}) \, ,
\label{bmod2}
\end{equation}
$C_{\perp}({\bf r})$ being a short-hand notation for
$C_{\perp\perp}({\bf r})$.
The chain bond follows from Eq.~(\ref{bond}),
or equivalently from the MSA-like
form of the correlation function, whose Fourier transform
is given by Eqs.~(\ref{oz}), (\ref{ck}). The transverse part is then
\begin{equation}
\hat{\Gamma}_{\perp}({\bf k})=\frac{R_{\perp}}
{1-R_{\perp}\left(c^{\perp}_{0}+qc^{\perp}_{1}\gamma({\bf k})\right)} \, ,
\label{gammamod}
\end{equation}
where $c^{\perp}_{0}$ can be regarded as representing the perturbing
interaction inside the hard core.
With $c^{\perp}_{1}=\beta$ as in the MSA, $R_{\perp}$ is clearly
the two-body hypervertex of the reference system.
With $c^{\perp}_{1}\neq \beta$, this interpretation can be kept
as the difference $c^{\perp}_{1}-\beta$ is small and will become even
smaller when including the new contribution $\Delta$. We then write
\begin{equation}
\hat{\Gamma}_{\perp}({\bf k})=
R_{\perp}+R_{\perp}^{2}\, \hat{C}_{\perp}({\bf k}) \, ,
\label{gammamod2}
\end{equation}
Along with Eq.~(\ref{gammamod}) this yields for the chain bond
in Fourier space
\begin{equation}
\hat{C}_{\perp}({\bf k})=\frac{z_{\perp}}{R_{\perp}P(z_{\perp})}\left[
\frac{\gamma({\bf k})}{1-z_{\perp}\gamma({\bf k})}
-\frac{P(z_{\perp})-1}{z_{\perp}}\right] \, .
\label{cbond}
\end{equation}
Note that since $\Gamma_{\perp}(r\!=\!0)=R_{\perp}$ we have
$C_{\perp}(r\!=\!0)=0$, which also follows from Eq.~(\ref{cbond}) using
the definition~(\ref{green}) for the integral $P(z)$.
Now the integral of the double chain bond can be evaluated
in a straightforward way by Fourier methods:
\begin{equation}
B=\int \! \! \frac{d^{3}{\bf k}}{(2\pi)^{3}} \,
\hat{C}_{\perp}^{2}({\bf k})=\left(\frac{1}
{R_{\perp} P(z_{\perp})}\right)^{2}
\left[z_{\perp} P'(z_{\perp})-P(z_{\perp})(P(z_{\perp})-1)\right] \, ,
\label{bmod}
\end{equation}
where we have set $P'(z)=dP(z)/dz$.
Inserting Eqs.~(\ref{vertex}) and~(\ref{bmod}) into Eq.~(\ref{delta}) yields
the sought quantity $\Delta$ to be used in Eq.~(\ref{ymod}).

\subsection{Spin-wave limit}

As $m\rightarrow 1$ the transverse modes represent harmonic oscillators
that can be solved exactly by Gaussian integrals. With the added contribution
$\Delta$ of Eq.~(\ref{chimod}) the SCOZA will recover the exact spin-wave
solution also for the longitudinal susceptibility as $m\rightarrow 1$.
I.e. thermodynamic self-consistency
is immediately fulfilled with $c^{\parallel}_{1}=c^{\perp}_{1}=\beta$ in
this limiting case also at phase equilibrium where
$P'(z_{\perp})\rightarrow \infty$.

To show that SCOZA has this behavior, we first evaluate $R_{\perp}$
as $m\rightarrow 1$. By setting $\nu=d/2$ we have
\begin{equation}
m=\langle s_{\bf r}^{\parallel}\rangle=
\langle\sqrt{1-d(s_{\bf r}^{\perp})^{2}}\rangle \sim 1-\nu
\langle (s_{\bf r}^{\perp})^{2}\rangle=1-\nu R_{\perp}
\label{mlow}
\end{equation}
so that in this limit
\begin{equation}
R_{\perp}=\frac{1-m}{\nu} \, .
\label{rlow}
\end{equation}
If we set $c^{\perp}_{1}=\beta$, Eq.~(\ref{c1}) then gives
\begin{equation}
q\, \beta(1-m)=\nu z_{\perp}P(z_{\perp}) \, .
\label{zlow}
\end{equation}
We also note that in the same limit $R_{\parallel}=
\langle (s_{\bf r}^{\parallel}-\langle s_{\bf r}^{\parallel}\rangle )^{2}
\rangle
=(1-m)^{2}/\nu$, so that it can be neglected compared with $R_{\perp}$.
Eqs.~(\ref{vertex}) and (\ref{rlow}) then imply
\begin{equation}
b=-\, \frac{(1-m)^{2}}{\nu^{2}} \, .
\label{vertlow}
\end{equation}
This together with Eqs.~(\ref{delta}), (\ref{bmod}), (\ref{zlow}) yields
\begin{equation}
\Delta=(1-m)^{2}\frac{1}{\nu P^{2}(z_{\perp})}
\left[z_{\perp}P'(z_{\perp})-P(z_{\perp})(P(z_{\perp})-1)\right] \, .
\label{deltalow}
\end{equation}
In order to obtain the longitudinal susceptibility from Eq.~(\ref{chimod})
one also needs knowledge of $y_{l}$. Eq.~(\ref{c1})
with $c^{\parallel}_{1}=\beta$
and $z=z_{l}$ implies $z_{l}\rightarrow q\beta R_{\parallel}
\rightarrow q\beta(1-m)^{2}/\nu$ (i.e. $P(z_{l})\rightarrow 1$), so that
Eq.~(\ref{y}) gives
\begin{equation}
y_{l} = 1-q\beta R_{\parallel} \rightarrow 1 \, .
\label{ylow}
\end{equation}
Substituting Eqs.~(\ref{deltalow}) and (\ref{ylow}) into Eq.~(\ref{chimod})
we obtain
\begin{equation}
\chi_{\parallel}= \frac{1}{\nu}(1-m)^{2}\,
\frac{z_{\perp}P'(z_{\perp})+P(z_{\perp})}{P^{2}(z_{\perp})}
\, .
\label{chilow}
\end{equation}
This can be compared with the result obtained by differentiating
the transverse susceptibility with respect to $m$. For $m\rightarrow 1$
only the transverse correlations contribute to the internal energy,
and the SCOZA equation~(\ref{consist}) reduces to
\begin{equation}
\frac{1}{q}\frac{\partial}{\partial \beta}\left(\frac{1}{\chi_{\perp}}\right)
=-1-\frac{\partial}{\partial m}\left(\nu R_{\perp}
\frac{P(z_{\perp})-1}{z_{\perp}P(z_{\perp})}\right)
=-1-\frac{\nu}{q\beta}P'(z_{\perp})\frac{\partial z_{\perp}}{\partial m} \, ,
\label{consistlow}
\end{equation}
where again Eq.~(\ref{c1}) with $c^{\perp}_{1}=\beta$ has been used.
In the same limit we then also have with Eqs.~(\ref{c1})
and (\ref{y})
\begin{equation}
y_{\perp}\rightarrow P(z_{\perp})-q\beta R_{\perp}
\label{yperplow}
\end{equation}
so that from Eqs.~(\ref{chi}) and (\ref{y})
\begin{equation}
\frac{1}{q}\frac{\partial}{\partial \beta}\left(\frac{1}{\chi_{\perp}}\right)
= -1+\frac{\nu}{q(1-m)}P'(z_{\perp})
\frac{\partial z_{\perp}}{\partial \beta} \, .
\label{consistlow2}
\end{equation}
The partial derivatives of $z_{\perp}$ which appear
in Eqs.~(\ref{consistlow}), (\ref{consistlow2}) are obtained
by differentiating Eq.~(\ref{zlow}) with respect to $\beta$ and $m$.
This gives
\begin{eqnarray}
\frac{\partial z_{\perp}}{\partial \beta} & = & \frac{q(1-m)}
{\nu [z_{\perp}P'(z_{\perp})+P(z_{\perp})]} \, ,
\label{zbetalow} \\
\frac{\partial z_{\perp}}{\partial m} & = & - \frac{q\beta}
{\nu [z_{\perp}P'(z_{\perp})+P(z_{\perp})]} \, .
\label{zmlow}
\end{eqnarray}
By substituting Eq.~(\ref{zbetalow}) into Eq.~(\ref{consistlow2})
it is readily seen that the consistency condition Eq.~(\ref{consistlow})
is satisfied. Finally by using Eqs.~(\ref{sum}), (\ref{chi}),
(\ref{y}), (\ref{yperplow}), (\ref{zmlow})
the longitudinal susceptibility becomes
\begin{equation}
\frac{1}{\chi_{\parallel}}=\frac{\partial}{\partial m}
\left(m\frac{y_{\perp}}{R_{\perp}}\right)\rightarrow
\nu \frac{\partial}{\partial m}
\left(\frac{P(z_{\perp})}{1-m}\right)\rightarrow \frac{\nu}{(1-m)^{2}} \,
\frac{P^{2}(z_{\perp})}{z_{\perp}P'(z_{\perp})+P(z_{\perp})} \, ,
\label{chilow2}
\end{equation}
in agreement with Eq.~(\ref{chilow}). Thus the contribution $\Delta$
in Eq.~(\ref{chimod}) fully accounts for the spin waves in the limit
$m\rightarrow 1$, and SCOZA
yields full thermodynamic self-consistency with
$c^{\parallel}_{1}=c^{\perp}_{1}=\beta$
in this limit. Without this term Eq.~(\ref{chilow}) reduces to
$\chi_{\parallel}=(1-m)^{2}/\nu=R_{\parallel}$ as a consequence of the
assumption $c^{\parallel}_{1}=\beta$ that leads
to Eq.~(\ref{deltalow}). However, SCOZA, that does not ``know'' about
this assumption, still requires Eq.~(\ref{chilow2}) to be fulfilled
(with $c^{\perp}_{1}\simeq \beta$) on the basis of the MSA-type
expression~(\ref{oz}), which implies
$\chi_{\parallel}=R_{\parallel}/[1-R_{\parallel}(c^{\parallel}_{0}
+qc^{\parallel}_{1})]$.
At phase equilibrium where $\chi_{\parallel}\rightarrow \infty$ since
$P'(z_{\perp})\rightarrow \infty$ this implies
that $c^{\parallel}_{0}+c^{\parallel}_{1}\rightarrow \infty$
as $R_{\parallel}\rightarrow 0$, so that $c^{\parallel}_{1}-\beta$
is no longer small and the longitudinal part of the correlation function
will be heavily distorted.

\section{Conclusions}

In this paper a simple version of our SCOZA is studied, in which the parallel
and transverse components of the onsite spin-spin correlation $R$ are assumed
to be those of the high-temperature susceptibility. This is found to be
quantitatively accurate in its assessment of $T_{c}$ but not in several other
respects. This treatment neglects the effects of the transverse correlations
on the longitudinal susceptibility. We then develop a version of SCOZA
in which the neglected correlations are included to lowest order in their
$\gamma$-ordered expansion. This guarantees that spin waves are well-accounted
for in a way that becomes exact in the $m\rightarrow 1$ limit.
The quantitative assessment of the resulting version of SCOZA will require
a significant
computational effort and is not attempted here. We defer it to our next paper
on the $D$-dimensional spin model.

\section*{Acknowledgments}

D.P. gratefully acknowledges the support of the Division of Chemical Sciences,
Office of Basic Energy Sciences, Office of Energy Research, U.S.~Department
of Energy for his work at Stony Brook. G.S. gratefully acknowledges
the support of the National Science Foundation.

\vspace*{3cm}

\begin{center}
TABLES
\end{center}

\begin{center}
\begin{tabular}{cccccc}  \hline \hline
\makebox[1.5cm]{lattice}   &
\makebox[1.5cm]{$\beta_{c,{\rm SCOZA}}$} &
\makebox[1.5cm]{$\beta_{c,{\rm QHRT}}$} &
\makebox[1.5cm]{$\beta_{c,{\rm ex}}$} &
\makebox[1.5cm]{$u_{c,{\rm SCOZA}}$} &
\makebox[1.5cm]{$u_{c,{\rm ex}}$} \\ \hline

\makebox[1.5cm]{SC}  & $0.6968$ & $0.7047$ & $0.6930^{a}$
& $-1.0216$ & $-0.987\pm 6^{c}$ \\
\makebox[1.5cm]{BCC} & $0.4887$ & & $0.4868^{a}$
& $-1.1289$ & $-1.092\pm 8^{c}$ \\
\makebox[1.5cm]{FCC} & $0.3160$ & & $0.3148^{b}$
& $-1.5379$ &\hspace{0.4cm}$-1.4808\pm 30^{c}$
\\    \hline \hline

\end{tabular}
\end{center}

\vspace*{1cm}

\noindent
Critical parameters of the classical Heisenberg model on a
SC, BCC, and FCC lattice. The inverse critical temperature $\beta_c$ and
the critical internal energy per spin $u_c$ predicted by the SCOZA
are compared with the ``exact'' $\beta_{c,{\rm ex}}$,
$u_{c,{\rm ex}}$ determined by extrapolation of high-temperature
expansions and by MC simulations. Energies and temperatures are in units
of the interaction strength $J$ and of $J/k_{\rm B}$ respectively.
{\em a}:~results from high-temperature expansions~\protect\cite{butera}
and MC-FSS simulations~\protect\cite{holm,landau}. {\em b}:~from
high-temperature expansions~\protect\cite{mckenzie}. {\em c}:~from
high-temperature expansions~\protect\cite{fisher}. For the SC lattice,
the inverse critical temperature predicted by the quantum hierarchical
reference theory (QHRT)~\protect\cite{gianinetti} in the classical limit
is also reported.

\newpage

\begin{figure}
\centerline{\psfig{file=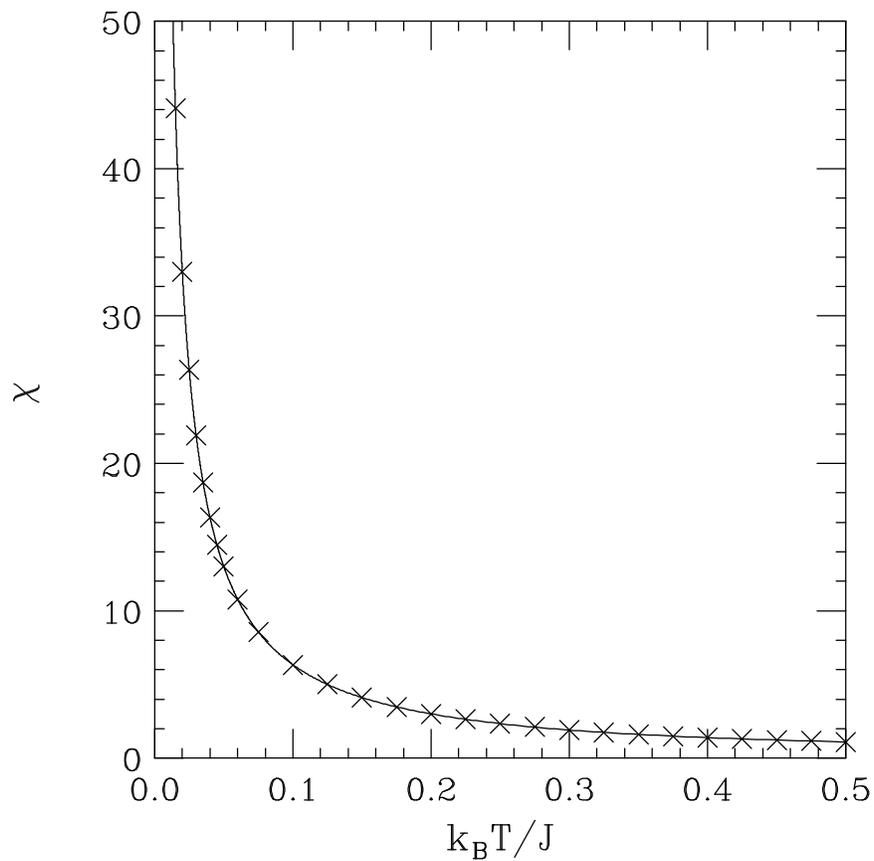,angle=90}}
\caption{
Reduced isothermal susceptibility $\chi$ in zero field
of the classical Heisenberg model on a one-dimensional chain as a function
of the dimensionless temperature $k_{\rm B}T/J$, $J$ being the spin-spin
coupling strength. Solid line:
SCOZA. Crosses: exact result~\protect\cite{onedim}.}
\end{figure}
\begin{figure}
\centerline{\psfig{file=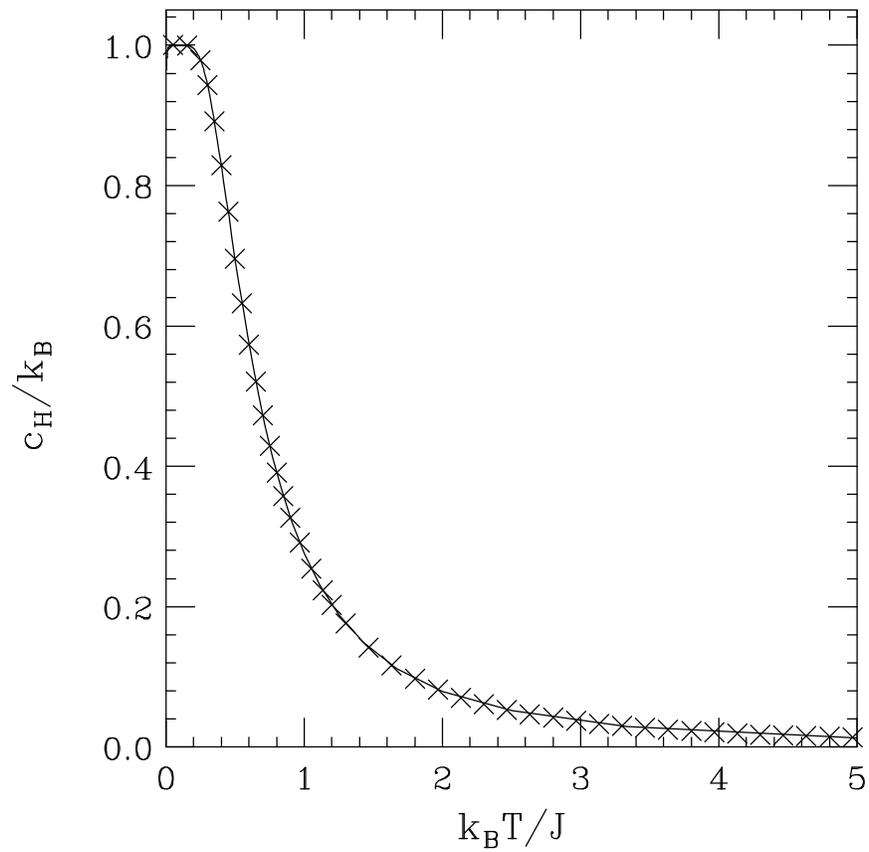,angle=90}}
\caption{Specific heat per spin $c_{H}$ in zero field of the classical
Heisenberg model on a one-dimensional chain as a function of the
dimensionless temperature $k_{\rm B}T/J$. Solid line: SCOZA. Crosses:
exact result~\protect\cite{onedim}.}
\end{figure}
\begin{figure}
\centerline{\psfig{file=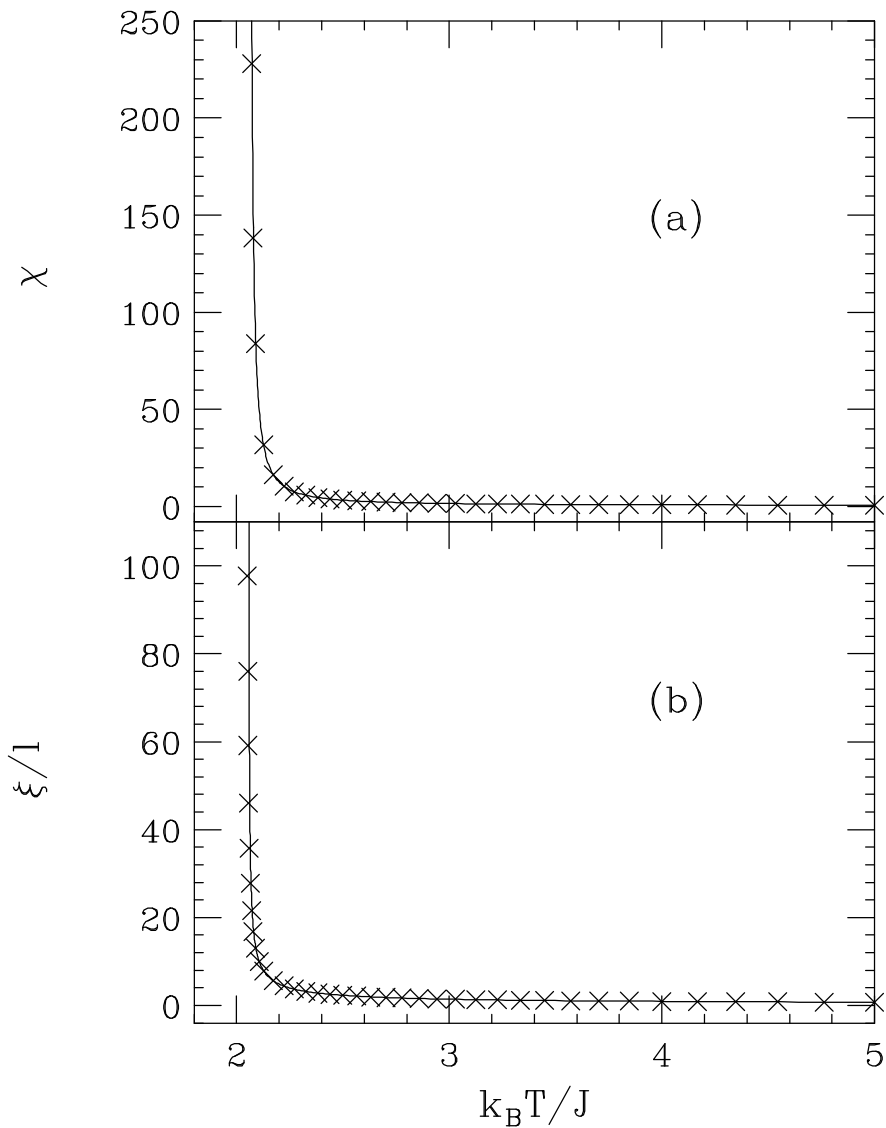,angle=180}}
\caption{
Reduced isothermal susceptibility $\chi$ (a) and correlation length
$\xi$ (b) of the classical Heisenberg model in zero field above the critical
temperature (BCC lattice). $\xi$ is in units of the lattice spacing $l$.
Crosses: SCOZA. Solid line: approximant~\protect\cite{fisher}.}
\end{figure}
\begin{figure}
\centerline{\psfig{file=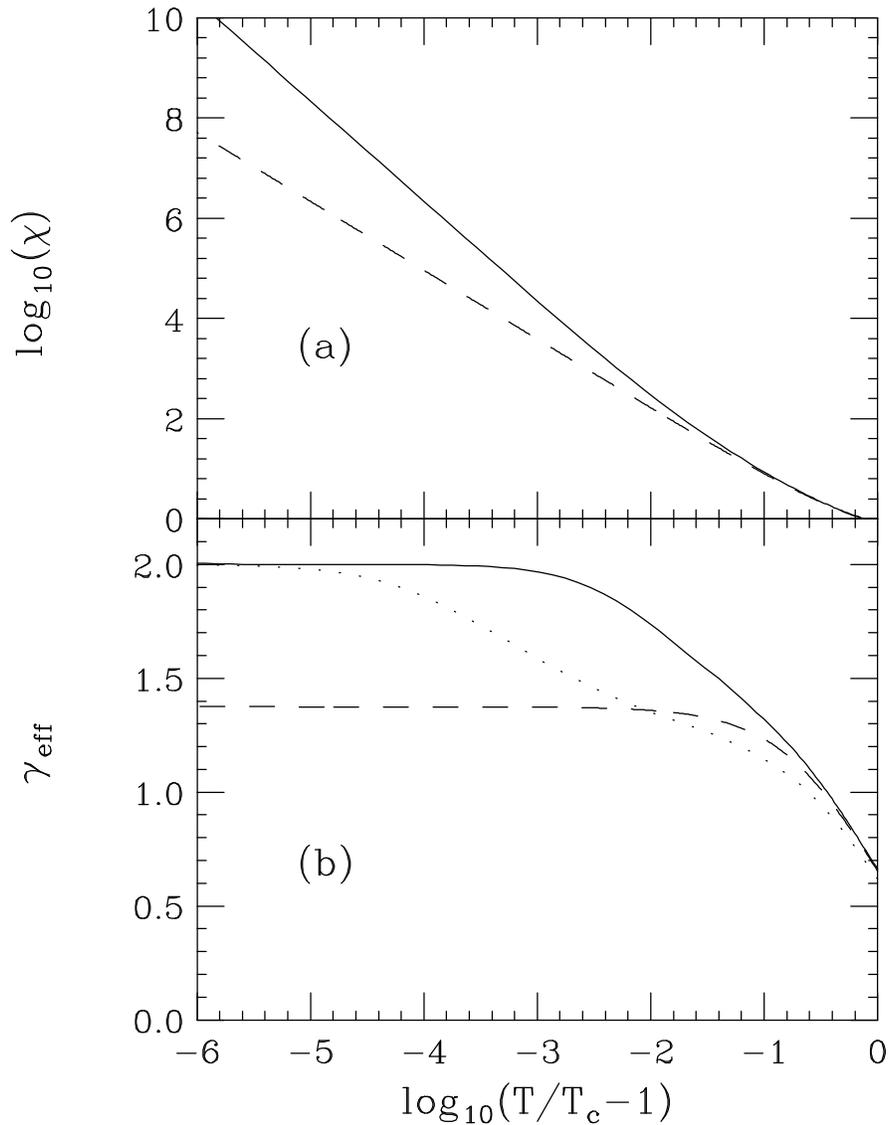,angle=180}}
\caption{
Reduced isothermal susceptibility $\chi$ in zero field (a)
and effective critical exponent $\gamma_{\rm eff}$ (b) of the
classical Heisenberg model (BCC lattice) as a function of the reduced
temperature $(T-T_c)/T_c$. Solid line: SCOZA. Dashed line:
approximant~\protect\cite{fisher}. Dotted line in panel (b): SCOZA result
for the Ising model~\protect\cite{ising}.}
\end{figure}
\begin{figure}
\centerline{\psfig{file=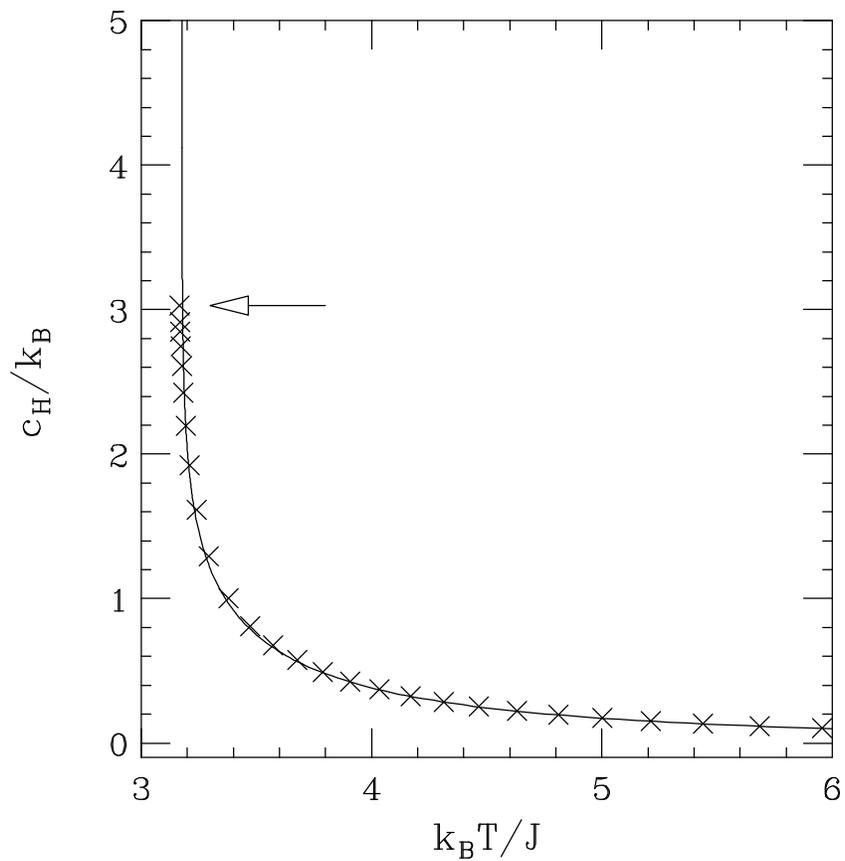,angle=90}}
\caption{
Specific heat per spin $c_{H}$ in zero field of the classical Heisenberg
model above the critical temperature (FCC lattice). Crosses: SCOZA.
Solid line: approximant~\protect\cite{domb}. The arrow marks the height
of the peak of the SCOZA specific heat $c_{\rm peak}\simeq 3.03$.}
\end{figure}
\begin{figure}
\centerline{\psfig{file=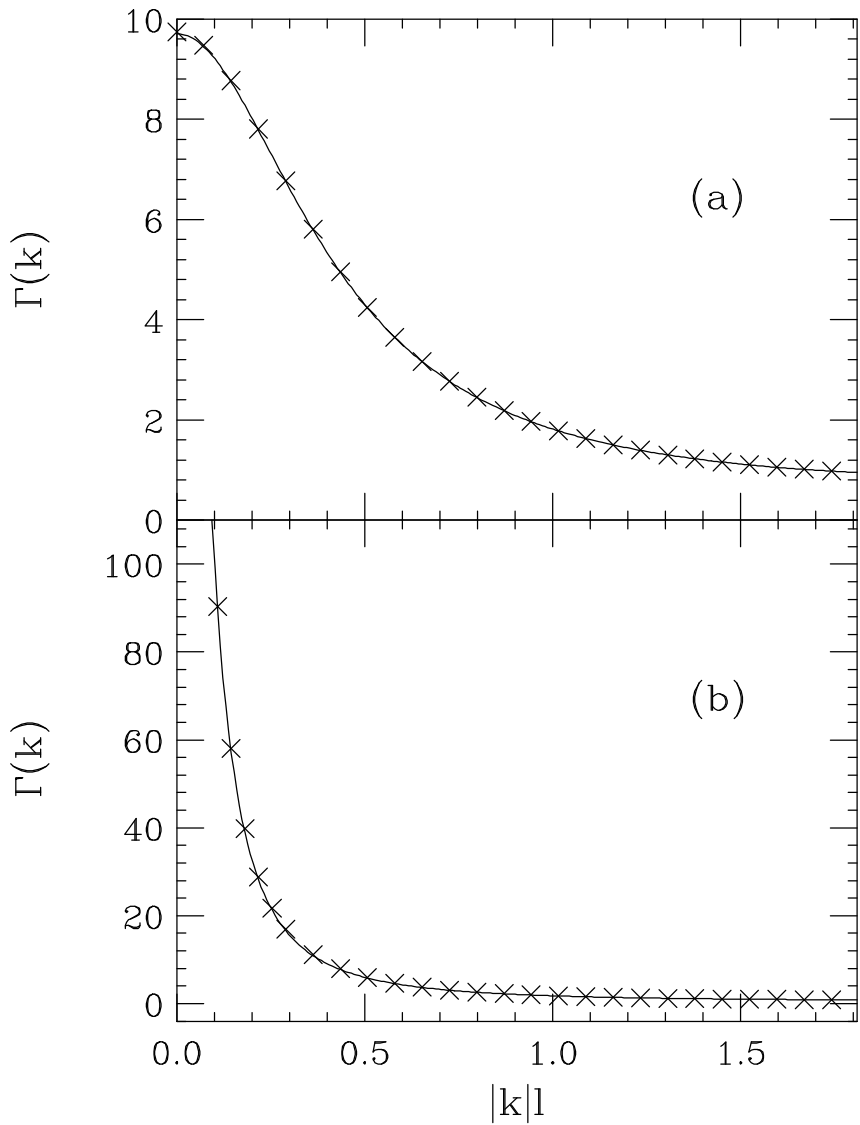,angle=180}}
\caption{
Magnetic structure factor $\Gamma({\bf k})$ in zero field of the
classical Heisenberg model on a BCC lattice along the direction
$k_x\!=k_y=\!k_z$. The norm of the wavevector $|{\bf k}|$ is in units of the
reciprocal of the lattice spacing $l$. The results shown are for inverse
temperature $\beta=0.4$ (a) and $\beta=0.48$ (b). Crosses: SCOZA. Solid line:
approximant~\protect\cite{fisher}.}
\end{figure}
\begin{figure}
\centerline{\psfig{file=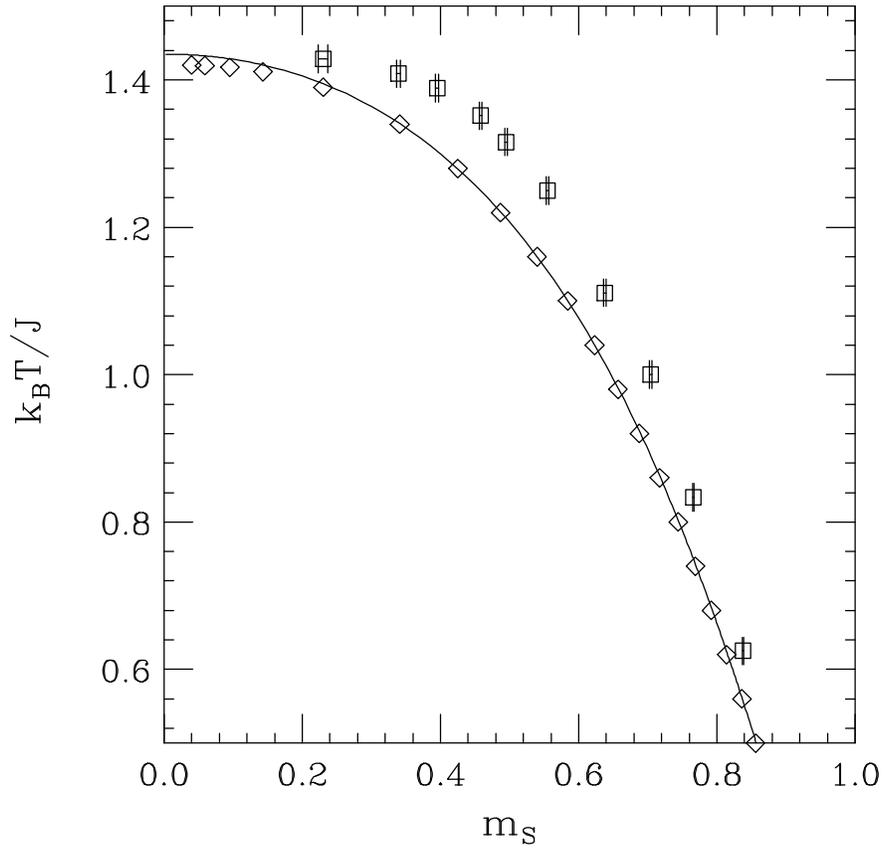,angle=90}}
\caption{
Spontaneous magnetization curve of the classical Heisenberg model
on a SC lattice. Solid line: SCOZA. Diamonds: QHRT~\protect\cite{gianinetti}.
Squares: MC simulation results~\protect\cite{binder}.}
\end{figure}
\begin{figure}
\centerline{\psfig{file=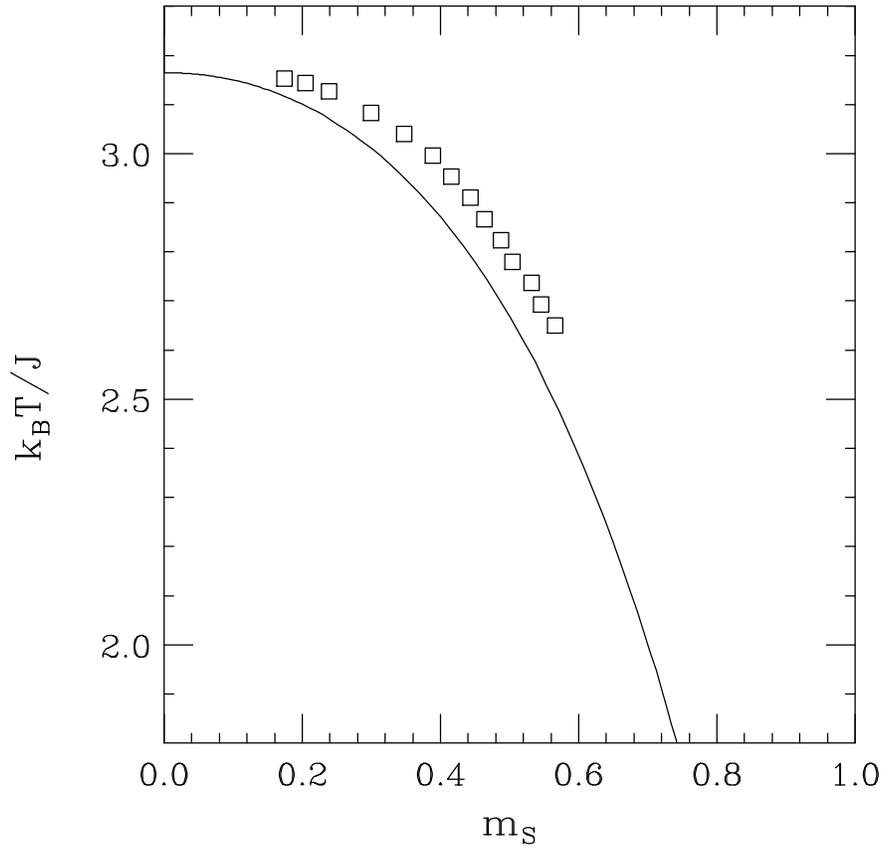,angle=90}}
\caption{
Spontaneous magnetization curve of the classical Heisenberg model
on a FCC lattice. Solid line: SCOZA. Squares: Pad\`e
approximant~\protect\cite{wood}.}
\end{figure}
\begin{figure}
\centerline{\psfig{file=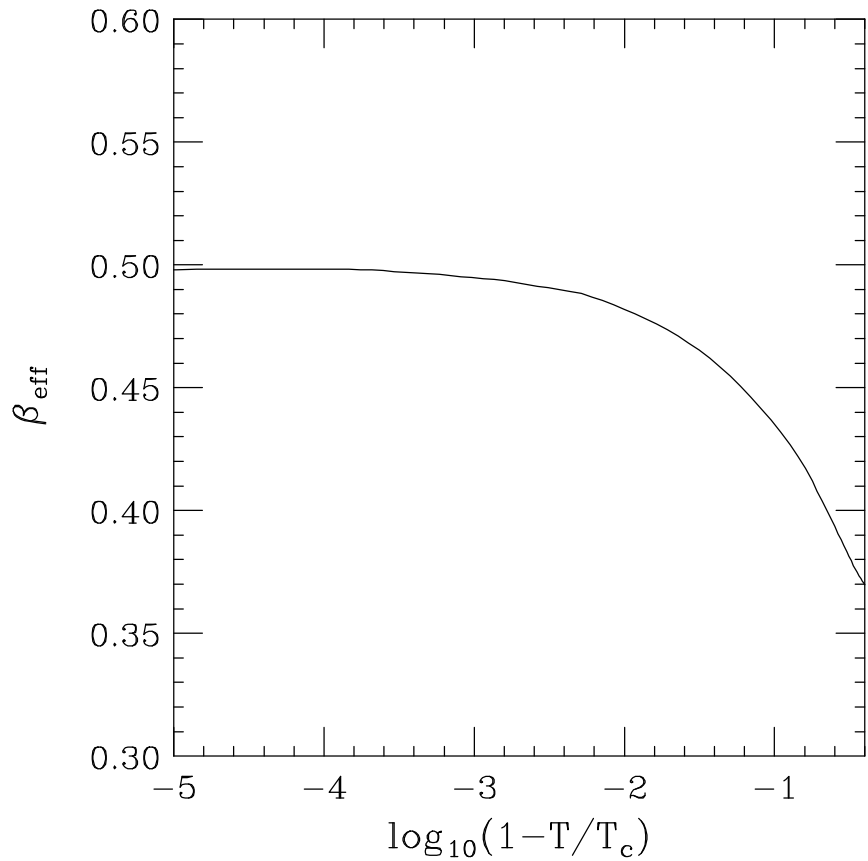,angle=90}}
\caption{
SCOZA effective exponent $\beta_{\rm eff}$ for the curvature of the
spontaneous magnetization curve of the classical Heisenberg model
(SC lattice) as a function of the reduced temperature $(T_c-T)/T_c$.}
\end{figure}
\begin{figure}
\centerline{\psfig{file=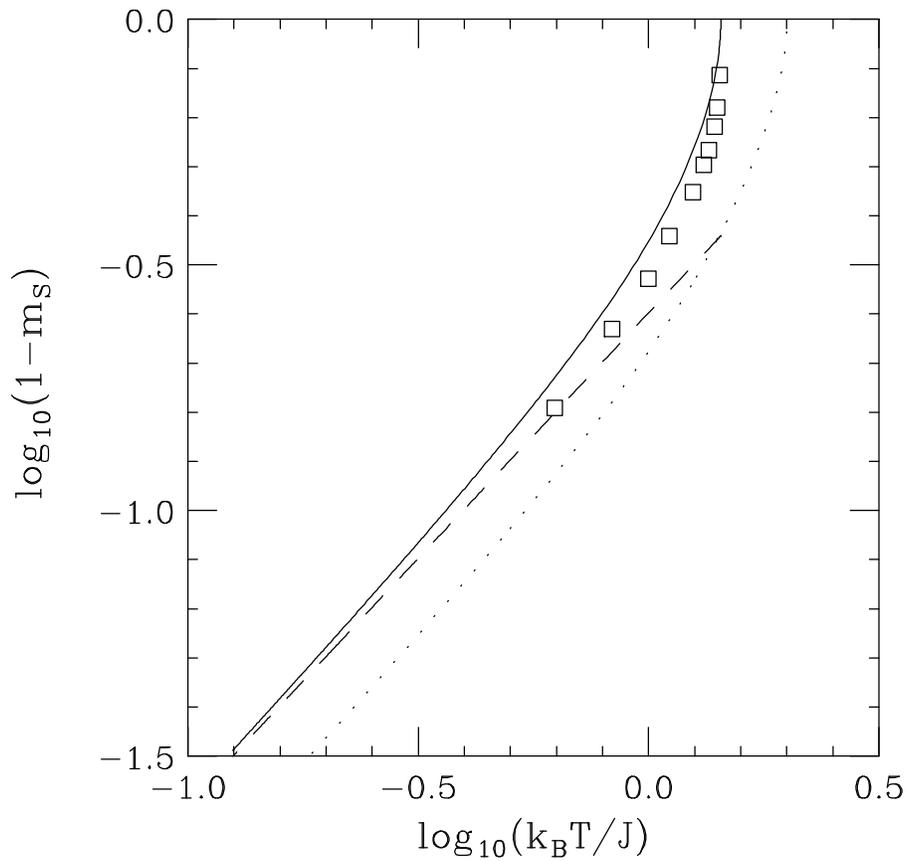,angle=90}}
\caption{
Low-temperature behavior of the spontaneous magnetization curve of
the classical Heisenberg model (SC lattice). Solid line: SCOZA. Dotted line:
mean-field theory. Squares: MC simulation data~\protect\cite{binder}.
Dashed line: spin-wave theory result to ${\cal O}(T)$~\protect\cite{spin}.}
\end{figure}
\begin{figure}
\centerline{\psfig{file=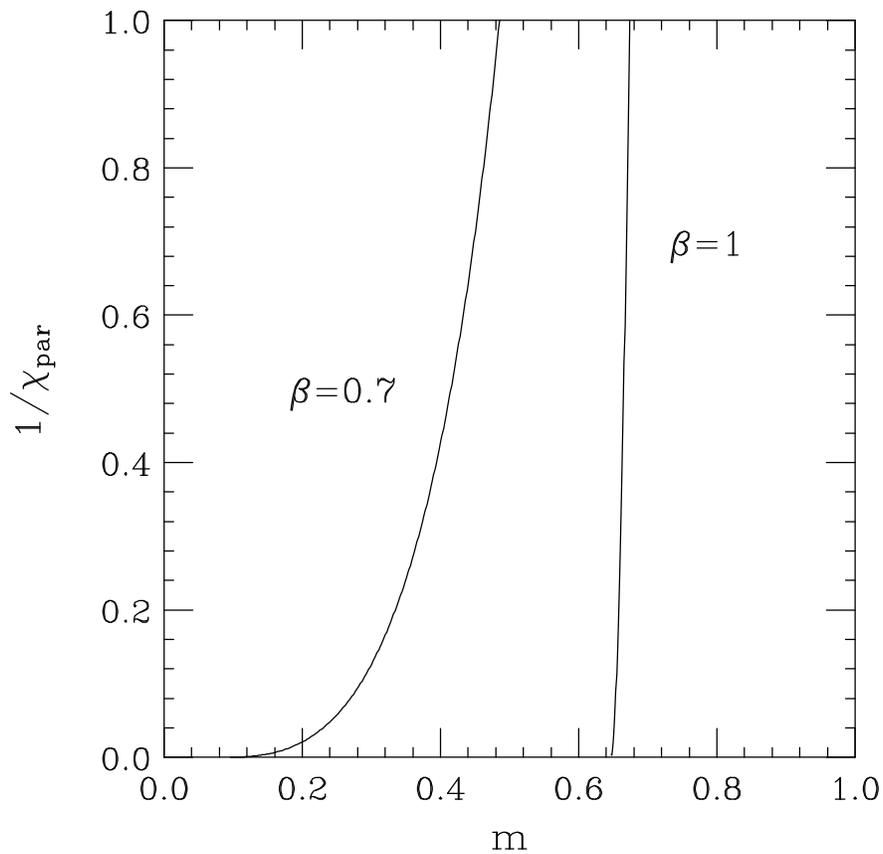,angle=90}}
\caption{
Reciprocal of the reduced longitudinal susceptibility
$1/\chi_{\parallel}$ of the classical Heisenberg model (SC lattice) outside
the phase boundary. The SCOZA results are shown for two different values
of the inverse temperature $\beta$.}
\end{figure}
\begin{figure}
\centerline{\psfig{file=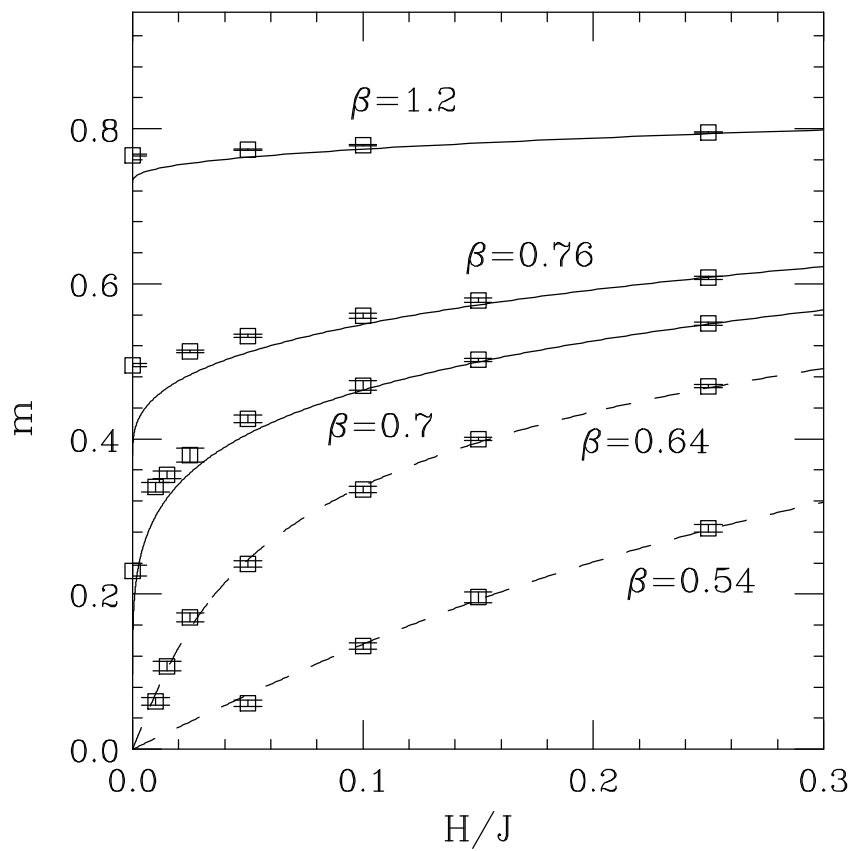,angle=90}}
\caption{
Magnetization isotherms of the classical Heisenberg model
(SC lattice). The magnetization $m$ is plotted as a function of the magnetic
field $H$ for the values of the inverse temperature $\beta$ shown
in the figure. Lines: SCOZA. Squares: MC results~\protect\cite{binder}.
Dashed and solid lines refer respectively to super- and sub-critical
isotherms.}
\end{figure}

\end{document}